\newif\ifonecol % Never comment out this line
\onecolfalse % Two-columns
\ifonecol
\else
\documentclass[letterpaper,conference]{IEEEtran}
\fi

\newif\ifsavespace
\savespacetrue

\ifonecol

\else
\fi
\usepackage{amsthm,amsmath,amssymb,color,graphicx,caption,subcaption,tikz,url,latexsym,xfrac,bm} %amsthm

\usepackage{verbatim}
\usetikzlibrary{decorations.markings}
\usetikzlibrary{shapes.geometric}
\usetikzlibrary{calc}
\usepackage{cite}
\usepackage{enumerate}
\usepackage{booktabs}
\usepackage{nicefrac}
\usepackage[draft,pdfborder={0 0 0}]{hyperref}

\usepackage{pgfplots}
\usepgfplotslibrary{units}
\usetikzlibrary{spy,backgrounds}
\usepackage{pgfplotstable}
\usepackage{flushend}
\usepackage[normalem]{ulem}
\usepackage{multirow}
\usepackage{glossaries} % For acronyms
\usepackage{pifont} %ding
\usepackage[commandnameprefix=ifneeded,commentmarkup=uwave,final]{changes}
\definechangesauthor[name=Pascal, color=blue]{P}
\definechangesauthor[name=Ilshat, color=purple]{I}
\definechangesauthor[name=Charles, color=red]{C}
\newcommand{\ubf}{\mathbf{u}}
\newcommand{\xbf}{\mathbf{x}}

\newcommand{\GN}{\mathbf{G}_N}
\newcommand{\Ical}{\mathcal{I}} % Information set
\newcommand{\Fcal}{\mathcal{F}} % frozen set
\newcommand{\Bcal}{\mathcal{B}} % Binary dominated set
\newcommand{\Bcalbf}{\bm{\mathcal{B}}} % Binary dominated set
\newcommand{\Mcal}{\mathcal{M}} % Partial dominated set
\newcommand{\Mcalbf}{\bm{\mathcal{M}}} % Partial dominated set

\newcommand{\quot}[1]{``#1''}
\newcommand{\Pmax}{P_{\text{max}}} % Max additional trials perturbation
\newcommand{\Fmax}{F_{\text{max}}} % Max additional trials flip
\newcommand{\Tmax}{T_{\text{max}}} % Total number of additional trials
\newcommand{\varp}{\sigma_{p}^2} % Variance of perturbation
\newcommand{\hatuvec}{\bm{\hat{u}}} %All bit-estimates 
\newcommand{\hatu}{\hat{u}} % for one bit estimate...
\newcommand{\Lvec}{\bm{\lambda}} %LLR vector 
\newcommand{\Lpvec}{\bm{\lambda}_p} % Corrupted LLR vector 
\newcommand{\adec}{\lambda_{\text{dec}}}
\newcommand{\HD}{\text{HD}}

\DeclareMathOperator{\SC}{SC}

\begin{document}
% set the "et al." parameters
\bstctlcite{IEEEexample:BSTcontrol}

\definecolor{darkred}{rgb}{0.75, 0, 0}
\definecolor{darkgreen}{rgb}{0, 0.45, 0}
\definecolor{bluegreen}{rgb}{0, 0.84, 0.84}
\definecolor{matlab1}{rgb}{0, 0.447, 0.741} % blue
\definecolor{matlab2}{rgb}{0.850, 0.325, 0.098} % red
\definecolor{matlab3}{rgb}{0.929, 0.694, 0.125} % yellow
\definecolor{matlab4}{rgb}{0.494, 0.184, 0.556} % violet
\definecolor{matlab5}{rgb}{0.466, 0.674, 0.188} % green
\definecolor{matlab6}{rgb}{0.301, 0.745, 0.933} % light blue
\definecolor{matlab7}{rgb}{0.635, 0.078, 0.184} % dark purple
\definecolor{ETSRed}{RGB}{239,62,69}
\newcommand{\mypoint}[2]{\tikz[remember picture]{\node[inner sep=0, anchor=base](#1){$#2$};}}
\usetikzlibrary{positioning}
%% Acronym definitions
\newacronym{scl}{SCL}{SC-List}
\newacronym{sc}{SC}{Successive-Cancellation}
\newacronym{ber}{BER}{Bit-Error-Rate}
\newacronym{bler}{BLER}{block-error rate}
\newacronym{llr}{LLR}{log-likelihood ratio}
\newacronym{awgn}{AWGN}{Additive White Gaussian Noise}
\newacronym{bpsk}{BPSK}{Binary Phase-Shift Keying}
\newacronym{ca}{CA}{CRC-aided}
\newacronym{crc}{CRC}{cyclic redundancy check}
\newacronym{db}{dB}{decibel}
\newacronym{snr}{SNR}{signal-to-noise ratio}
\newacronym{scfp}{SCFP}{Successive-Cancellation Flip and Perturbation}
%\newacronym{dscfp}{DSCFP}{Dynamic SCFP}
\newacronym{dscfp}{DSCFP}{DSCF and Perturbation}
\newacronym{scf}{SCF}{SC Flip}
\newacronym{scp}{SCP}{SC Perturbation}
\newacronym{dscf_abstract}{DSCF}{Dynamic SC Flip}
\newacronym{dscf}{DSCF}{Dynamic SCF}
\newacronym{pdscf}{PDSCF}{Perturbed DSCF}
\newacronym{dscp}{DSCP}{Dynamic SCP}
\newacronym{cnn}{CNN}{Convolutional Neural Network}
\newacronym{adscfp}{ADSCFP}{Accumulative DSCFP}
\newacronym{apdscf}{APDSCF}{Accumulative PDSCF}
%----------------------------------
\title{Successive-Cancellation Flip and Perturbation Decoder of Polar Codes}
%\author{Author 1, Author 2, Author 3, Author 4}
\author{\IEEEauthorblockN{Charles Pillet, Ilshat Sagitov, Dominic Deslandes,  and Pascal Giard}

%\thanks{Charles Pillet, Ilshat Sagitov, and Pascal Giard (\{charles.pillet, ilshat.sagitov\}@lacime.etsmtl.ca, pascal.giard@etsmtl.ca) are with LaCIME, École de technologie supérieure (ÉTS), Montréal, QC, Canada. Dominic Deslandes (dominic.deslandes@sparkmicro.com) is with SPARK Microsystems.}}

  \IEEEauthorblockA{Department of Electrical Engineering, \'Ecole de technologie sup\'erieure, Montr\'eal, Qu\'ebec, Canada.\\Email: \{charles.pillet.1,ilshat.sagitov.1\}@ens.etsmtl.ca,  \{dominic.deslandes,pascal.giard\}@etsmtl.ca}%
}
\maketitle
 
 \begin{abstract}
    In this paper, two decoding algorithms based on \gls{sc} are proposed to improve the error-correction performance of \gls{crc}-aided polar codes  while aiming for a low-complexity implementation.
    Comparisons with \gls{dscf_abstract} and \gls{scp} are carried out since the proposed 
    \gls{dscfp} and \gls{pdscf} algorithms combine both methods. % to decode a noisy vector.
    The analysis includes comparisons with several code lengths $N$ and various number of decoding attempts $T_{max}$. 
    For $N=1024$ and the coding rate $R=\nicefrac{1}{2}$, the \gls{dscf_abstract} and the \gls{scp} algorithms with $\Tmax=17$ are bested by approximately $0.1$\,dB at     \gls{bler} of $0.001$. 
    At $\text{BLER}=10^{-6}$ and for $\Tmax=64$, the gain is of $0.375$\,dB and $>0.5$\,dB with respect to \gls{dscf_abstract} and \gls{scp}, respectively.
    At high signal-to-noise ratio, the average computational complexity of the proposed algorithms is virtually equivalent to that of \gls{sc}.
    
\end{abstract}
%% Good practices dictate that we reset the acronyms' definitions after the abstract
\glsresetall
	\IEEEpeerreviewmaketitle	%===========================================================================================
 
\section{Introduction}
    Polar codes are proved to be capacity-achieving on binary memoryless channels under low-complexity \gls{sc} decoding \cite{arik_polariz}, when the code length $N$ goes to infinity.
    Polar codes concatenated with a \gls{crc} code, i.e., \gls{ca}-polar codes are standardized in 5G \cite{3GPP_5G_Coding}, with code lengths $N\leq1024$.
    For these lengths, \gls{sc} error-correction capability is substandard such that \gls{scl} decoding \cite{scl_intro} is used as baseline for the 5G error-correction performance evaluation.
    
    \gls{scl} returns $L$ distinct codeword candidates at the end of decoding.
    The most probable candidate matching the \gls{crc} code is chosen as decoding output.
    \Gls{scl} implementations typically require $L$ times more memory and processing elements with respect to \gls{sc} \cite{fast_sscl}.
    %Moreover, an advanced sorting module is needed.
    In average, doubling the list size $L$ doubles the required area \cite{fast_sscl,pract_dscf}.
        
    Unlike \gls{scl}, \gls{scf} \cite{scf_intro} reuses the same \gls{sc} instance and applies one bit-flip during an additional trial.
    Decoding stops when the \gls{crc} code is checked by one of the estimated candidates or after $\Fmax$ additional trials.
    With a small $\Fmax$, error-correction performance heavily depends on the list of bit-flipping candidates. 
    The construction of that list has been improved with \gls{dscf}, where a better metric is proposed \cite{dyn_scf} while supporting multiple bit flips.
    Multi-bit flip \gls{dscf} improves error-correction performance but it increases memory requirements and therefore the required area of an implementation \cite{pract_dscf}. Furthermore, that coding gain also comes at the cost of an increased latency \cite{dyn_scf}.
    Meanwhile, single-bit flip decoder incurs a negligible hardware overhead \cite{Giard_JETCAS_2017}.

    \gls{scp} \cite{SCP} shares several properties with the flip decoding algorithm \cite{scf_intro,dyn_scf}.
    Most notably, \gls{scp} also requires a single \gls{sc} decoder instance.
    During an additional trial, the vector of \glspl{llr} is corrupted by Gaussian noise with an intensity either chosen manually \cite{SCP}, dynamically \cite{dyn_SCP}, with a \gls{cnn} \cite{CNN_perb}, or neurally adjusted \cite{neur_adjusted}.
    The last three involve either more complex circuitry, additional trials, or training to improve the error-correction capability.
    The elementary \gls{scp} decoding algorithm \cite{SCP} is shown to have an error-correction performances similar to \gls{scf} \cite{SCF_SCP_COMP}.%, but the single bit-flip \gls{dscf} was omitted in this work.
    %\gls{dscf} remains the algorithm with the best error-correction performance for an area similar to \gls{sc}.

    In this paper, two elementary decoding algorithms based on reusing the same \gls{sc} instance are proposed: \gls{dscfp}, and \gls{pdscf}. Both algorithms combine perturbation and flip decoding algorithms to improve error-correction performance with respect to \gls{scp} and \gls{dscf}. % for various maximum trials $\Tmax$. % for small $\Tmax$.
    \Gls{dscfp} exhibits better performance for small maximum number of trials $\Tmax$ while \gls{pdscf} exhibits better performance for larger $\Tmax$.
    For the code lengths $N=\{512,1024\}$ and the code rate $\nicefrac{1}{2}$, \gls{dscfp} is on par with \gls{dscf} at a \gls{bler} of $10^{-3}$ and $\Tmax=9$ while offering a gain of approximately $0.2$\,dB over \gls{scp}.
    For $\Tmax=17$, the gain is around $0.1$\,dB over both \gls{scp} and \gls{dscf}.
    For $\Tmax=33$, the gain is of approximately $0.2$\,dB and $0.1$\,dB over \gls{dscf} and \gls{scp}, respectively.   
    At $\text{\gls{bler}}=10^{-6}$ and $\Tmax=64$, \gls{pdscf} shows $0.375$\,dB and more than $0.5$\,dB gain with respect to \gls{dscf} and \gls{scp}.
    For $\text{BLER}<10^{-2}$, both proposed algorithms are shown to have an average computational complexity similar to that of \gls{sc}.

\section{Preliminaries}
    \subsection{Polar Codes and Successive Cancellation Decoding}
    A polar code \cite{arik_polariz}  of length $N=2^n$ and dimension $K$, denoted as $(N,K)$, is fully defined by its information set $\Ical\subseteq\{0,\dots,N-1\}$ with $|\Ical|=K$. 
    The information set is usually composed of the $K$ most reliable virtual channels resulting from the polarization induced by the transformation matrix $\GN=\left[\begin{smallmatrix}
        1&0\\1&1
    \end{smallmatrix}\right]^{\otimes n}$ \cite{arik_polariz}.
    The other $N-K$ virtual channels compose the frozen set $\Fcal=\Ical^c$.
    A codeword $\xbf=\{x_0,\dots,x_{N-1}\}$ of the polar code is computed as $\xbf=\ubf\GN$ where $\ubf=\{u_0,\dots,u_{N-1}\}$ is a vector of $N$ bits in which the message is stored in $\Ical$, i.e., $u_i=\{0,1\}$ if $i\in\Ical$ while the other positions are frozen to 0, i.e., $u_i=0$ if $i\in\Fcal$.
    A \gls{crc} code of $C$ bits is able to detect an error when providing a candidate $\hatuvec$ for $\ubf$.
    The additional $C$ \gls{crc} bits are as well stored in $\ubf$, increasing the size of $\Ical$ from $K$ to $K+C$ elements.
    Next, the notation $\Ical=\{i_0,\dots,i_j,\dots,i_{K+C-1}\}$ is used to describe the elements of $\Ical$, it verifies $i_0<\dots<i_j<\dots<i_{K+C-1}$.
    %This simple concatenated scheme is referred as a \gls{crc}-aided (CA)-polar code.

    Polar codes have been proposed with the low-complexity \gls{sc} algorithm.
    \gls{sc} is a tree decoding  algorithm in which the bit estimate vector $\hatuvec=\{\hatu_0,\dots,\hatu_{N-1}\}$ is obtained sequentially on the basis of the \gls{llr} vector $\Lvec$ and the information set $\Ical$.
    In more detail, for each of the $N$-leaves, a decision \gls{llr} $\adec$ is computed and a bit estimate is taken according to $\adec$ and $\Ical$. 
    Namely, at the $i^{th}$ leaf, $\hatu_i$ is computed as
    \begin{align}
        \hatu_i=\begin{cases}
        0\,, \quad &\text{if} \;\; i\in\Fcal, \\
        \HD(\adec(i))\,, \quad \;\;\,&\text{if} \;\; i\in\Ical,
        \end{cases}\label{eq:decision}
    \end{align}
    where $\HD\left(\adec(i)\right)$ corresponds to the hard decision on the decision \gls{llr} $\adec(i)$ at the $i^{th}$ leaf.
    Next, the notation $\hatuvec=\SC(\Lvec)$ is used to denote a decoding attempt with \gls{sc} algorithm fed with the vector of \glspl{llr} $\Lvec$.
    Note that the \gls{sc} decoding algorithm does not take advantage of the \gls{crc} code.
    In the remainder of the paper, CA-polar codes are used with the notation $(N,K+C)$.
    \subsection{Flip Algorithm}
    A decoding algorithm taking advantage of the \gls{crc} is the flip algorithm of polar codes.
    The \gls{scf} \cite{scf_intro} and \gls{dscf} \cite{dyn_scf} decoding algorithms are characterized by the parameter $\Fmax$, defining the maximum number of \gls{sc} additional trials.
    An additional trial is performed whenever the estimation $\hatuvec$ of the previous \gls{sc} trial fails to check the \gls{crc} code leading to $\hatuvec\neq\ubf$.
    During the initial \gls{sc} trial, $K+C$ metrics are calculated, one per information bit.
    The set of metrics is denoted $\Mcalbf=\{\Mcal_{i_0},\dots,\Mcal_{i_j},\dots,\Mcal_{i_{K+C-1}}\}$ where $\Mcal_{i_j}$ represents the metric for the $j^{th}$ information bit, being $i_j$ by definition.
    \Gls{scf} and \gls{dscf} have a different metric. 
    In the scenario where a single bit-flip is allowed for \gls{dscf}, the metrics are approximated as
    \begin{align}
        \Mcal_j = \begin{cases}
            |\adec(i_j)|, &\text{for SCF},\\
            |\adec(i_j)|+\sum_{k=0}^j J(\adec(i_k)),  \quad&\text{for DSCF},
        \end{cases}\label{eq:metric}
    \end{align}
    where $J(\cdot)$ is an approximation function \cite{simp_dscf} with
    \begin{align}
        J(i_k) = \begin{cases}
        1.5\,, \quad \text{if} \;\; \left|\adec(i_k)\right|\leq 5.0, \\
        0\,, \quad \;\;\,\text{otherwise}.
        \end{cases}
    \end{align}
    The metric of \gls{dscf} includes an accumulative part leading to a more reliable set of bit-flipping candidates.
    This set, denoted as $\Bcalbf=\{\Bcal(1),\dots,\Bcal(\Fmax)\}$, gathers the information bit locations of the $\Fmax$ smallest metrics in $\Mcalbf$.
    The  $t^{th}$ additional trial is characterized by the bit-flipping location $\Bcal(t)$ where the reversed bit decision with respect to the initial trial is taken.
    Namely, with respect to \eqref{eq:decision}, we have
    \begin{align}
        \hatu_i=\begin{cases}
        0\,, \quad &\text{if} \;\; i\in\Fcal, \\
        \HD(\adec(i))\,, \quad \;\;\,&\text{if} \;\; i\in\Ical,\\
        \HD(\adec(i))\oplus1\,, \quad \;\;\,&\text{if} \;\; i=\Bcal(t).
        \end{cases}
    \end{align}
    In the remainder of the paper, the notation $\hatuvec=\SC(\Lvec,\Bcal(t))$ describes the  $t^{th}$ \gls{sc} additional trial with $\Bcal(t)$ being the bit-flipping candidate and $\Lvec$ being the vector of \glspl{llr}.
    Moreover, the \gls{dscf} decoder fed by $\Lvec$ and returning $\hatuvec$ is noted $\hatuvec=\text{DSCF}(\Lvec)$.
    In order to reach the error-correction saturation, the reliable metric of \gls{dscf} reduces the required value of $\Fmax$ with respect to \gls{scf}. 
    %while having greater error-correction capability.
    %The performance of \gls{dscf} saturates for a small value of $\Fmax$.
    Note that \gls{dscf} may allow more than one bit-flip locations, where common values are 2 and 3, improving the error-correction performance even more.
    However, it comes at the cost of an increased $\Fmax$ to reach the error-correction saturation, respectively, $\times 5$ and $\times 30$ \cite{dyn_scf}.
    Moreover, $25\%$ and $81\%$ area increase is required to implement \gls{dscf} with 2 and 3 bit-flipping candidates per additional trial \cite{pract_dscf}, while the implementation required specific nodes to handle 2 or 3 bit-flipping candidates.
    \gls{dscf} with a single bit flip has an area close to \gls{sc} and its implementation is less complicated \cite{pract_dscf,fast-SCF}.
    
    \subsection{Perturbation Algorithm}
    %In contrary to flip decoding algorithm, the perturbation decoder has not been extensively studied in the literature, despite showing similarities with the flip decoding.
    Another algorithm taking advantage of the \gls{crc} code is the \gls{scp} decoding algorithm \cite{SCP}.
    As for \gls{dscf}, \gls{scp} exhibits a variable execution time.
    \gls{scp} successively tries the \gls{sc} algorithm until its estimation $\hatuvec$ checks the \gls{crc} code.
    Up to $\Pmax$ \gls{sc} additional trials  fed with a corrupted \gls{llr} vector $\Lpvec$ are tried, i.e., the additional trial corresponds to $\hatuvec=\text{SC}(\Lpvec)$.
    For any of the additional trials of the elementary \gls{scp} algorithm \cite{SCP}, the corrupted \gls{llr} vector is computed as 
    \begin{align}
        \Lpvec = \Lvec+\mathcal{N}(0,\varp),\,\label{eq:noisesimple}
    \end{align} 
    where $\varp$ is the variance of the Gaussian perturbation added to the LLR vector. 
    In \cite{dyn_SCP}, the authors proposed the \gls{dscp} decoding algorithm, in which $\varp$ is successively increased by a constant if the current trial returns a codeword already estimated by one previous trial.
    The performance is slightly improved with respect to \cite{SCP}, especially for small $\Pmax$. 
    However, the values of the initial $\varp$ and of the variance step need to be studied prior to the decoding.
    Moreover, it comes at the cost of storing $\Pmax+1$ estimated codewords, increasing the memory requirements, and thus the area.

    %In \cite{neur_adjusted}, the \gls{scp} decoder is first tried using \eqref{eq:noisesimple}. 
    %If all $\Pmax$ additional trials failed, a directed neural-evolutionary noise is generated and added to $\Lvec$, which is then fed to \gls{sc}.
    %By adding $10\Pmax$ more additional trials, around $0.5$\,dB gain is observed with respect to \gls{scp}.
    %Flip and perturbation decoders share several properties.
    %They have a variable execution time and are based on successive trials of the \gls{sc} decoding algorithm.
    Error-correction performance comparison has been performed between \gls{scp} \cite{SCP} and \gls{scf} \cite{scf_intro} in \cite{SCF_SCP_COMP}, no comparison is performed for the more reliable \gls{dscf} decoder with a single bit-flip.
    The comparison is limited to polar codes with $N\leq512$.
    It is shown that for small $\Pmax=\Fmax\leq15$, SCF has better error-correction capability.
    However, the performance saturates for \gls{scf} such that \gls{scp} exhibits greater error-correction performances for $\Pmax=\Fmax=64$.
        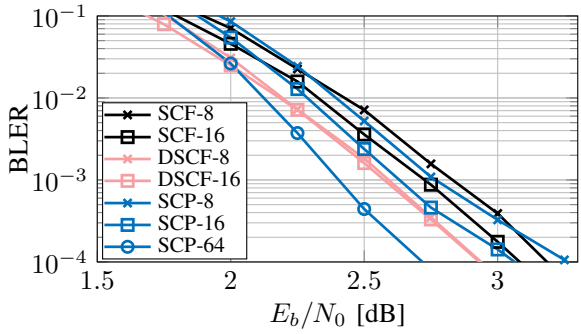
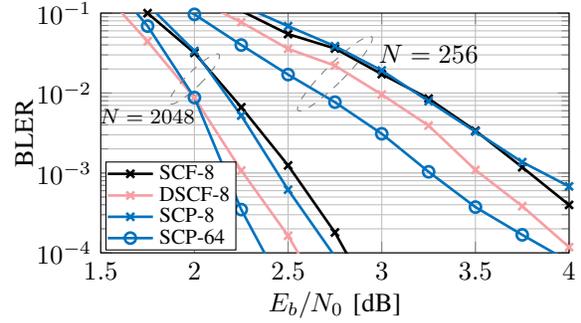
\begin{figure*}[ht]
      	\begin{subfigure}{0.99\columnwidth}
	       \centering
            \resizebox{0.9\columnwidth}{!}{\begin{tikzpicture}
  \pgfplotsset{
    label style = {font=\fontsize{10pt}{8.2}\selectfont},
    tick label style = {font=\fontsize{10pt}{8.2}\selectfont}
  }
  \begin{semilogyaxis}[
    width=0.9\columnwidth,
    height=0.55\columnwidth,
    xmin=1.5, xmax=3.3,
    xlabel={$E_b/N_0$ [dB]},
    xlabel style={yshift=0.4em},
    ymin=0.0001, ymax=0.1,
    ylabel style={yshift=-0.6em},
    ylabel={BLER},
    yminorticks, xmajorgrids,
    ymajorgrids, yminorgrids,
    %legend pos=outer east,
    legend style={at={(0.01,0.01)},anchor=south west},
    %column font, row sep
    legend style={legend columns=1, font=\footnotesize, row sep=-1mm},
    legend style={fill=white, fill opacity=1, draw opacity=1,text opacity=1}, % for future use maybe ? %opacity of filling/border and inside text
    legend style={inner xsep=0.2pt, inner ysep=-1pt}, % TIGHTER
     legend cell align={left},
    mark size=1.6pt, mark options=solid,
    ]      
            \addplot[color=black, mark=x, line width=1pt, mark size=2.1pt]
table[x=Eb,y=FER]{figures/data_SCF/1024_R12_16_8.txt}; 
\addlegendentry{SCF-8}
        \addplot[color=black,  mark=square, line width=1pt, mark size=2.1pt]
table[x=Eb,y=FER]{figures/data_SCF/1024_R12_16_16.txt}; 
    \addlegendentry{SCF-16}

\addplot[color=ETSRed!50,   mark=x, line width=1pt, mark size=2.1pt]
table[x=Eb,y=FER]{figures/data_DSCF/1024_R12_16_8.txt}; 
    \addlegendentry{DSCF-8}
\addplot[color=ETSRed!50,  mark=square, line width=1pt, mark size=2.1pt]
table[x=Eb,y=FER]{figures/data_DSCF/1024_R12_16_16.txt}; 
    \addlegendentry{DSCF-16}

   \addplot[color=matlab1!100,   mark=x, line width=1pt, mark size=2.1pt]
table[x=Eb,y=FER]{figures/data_SCP/1024_R12_16_8.txt}; 
    \addlegendentry{SCP-8}
\addplot[color=matlab1!100,  mark=square, line width=1pt, mark size=2.1pt]
table[x=Eb,y=FER]{figures/data_SCP/1024_R12_16_16.txt}; 
    \addlegendentry{SCP-16}

    \addplot[color=matlab1!100,  mark=o, line width=1pt, mark size=2.1pt]
table[x=Eb,y=FER]{figures/data_SCP/1024_R12_16_64.txt}; 
    \addlegendentry{SCP-64}

 %       \addplot[color=matlab1!50,   mark=x, line width=1pt, mark size=2.1pt]
%table[x=Eb,y=FER]{figures/data_DSCP/1024_R12_10.txt}; 
%    \addlegendentry{DSCP-8}
%\addplot[color=matlab1!50,  mark=square, line width=1pt, mark size=2.1pt]
%table[x=Eb,y=FER]{figures/data_DSCP/1024_R12_20.txt}; 
%    \addlegendentry{DSCP-16}
%    \draw[<->,ETSETSRed] (axis cs:1.45,0.1) -- (axis cs:3.875,0.1) node [midway,above] {2.42 dB};

 %   \draw[<->,black] (axis cs:3.875,0.1) -- (axis cs:7.9,0.1)  node [midway,below] {\large $\Delta\nicefrac{E_b}{N_0}=4$ dB} node [midway,above] {\large $\Delta\nicefrac{E_s}{N_0}=7.17$ dB};
  \end{semilogyaxis}

\end{tikzpicture}}%\vspace{-6pt}
            \caption{}
            \label{fig:comp_SCF_SCP_1024}
        \end{subfigure}
        \begin{subfigure}{0.99\columnwidth}
	       \centering
            \resizebox{0.9\columnwidth}{!}{\begin{tikzpicture}
  \pgfplotsset{
    label style = {font=\fontsize{10pt}{8.2}\selectfont},
    tick label style = {font=\fontsize{10pt}{8.2}\selectfont}
  }
  \begin{semilogyaxis}[
    width=0.9\columnwidth,
    height=0.55\columnwidth,
    xmin=1.5, xmax=4,
    xlabel={$E_b/N_0$ [dB]},
    xlabel style={yshift=0.4em},
    ymin=0.0001, ymax=0.1,
    ylabel style={yshift=-0.6em},
    ylabel={BLER},
    yminorticks, xmajorgrids,
    ymajorgrids, yminorgrids,
    %legend pos=outer east,
    legend style={at={(0.01,0.01)},anchor=south west},
    %column font, row sep
    legend style={legend columns=1, font=\footnotesize, row sep=-1mm},
    legend style={fill=white, fill opacity=0.8, draw opacity=1,text opacity=1}, % for future use maybe ? %opacity of filling/border and inside text
    legend style={inner xsep=0.2pt, inner ysep=-1pt}, % TIGHTER
     legend cell align={left},
    mark size=1.6pt, mark options=solid,
    ]

\addplot[color=black, mark=x, line width=1pt, mark size=2.1pt]
table[x=Eb,y=FER]{figures/data_SCF/256_R12_16_8.txt}; 
\addlegendentry{SCF-8}

\addplot[color=ETSRed!50,   mark=x, line width=1pt, mark size=2.1pt]
table[x=Eb,y=FER]{figures/data_DSCF/256_R12_16_8.txt}; 
\addlegendentry{DSCF-8}

\addplot[color=matlab1!100,   mark=x, line width=1pt, mark size=2.1pt]
table[x=Eb,y=FER]{figures/data_SCP/256_R12_16_8.txt}; 
\addlegendentry{SCP-8}
\addplot[color=matlab1!100,   mark=o, line width=1pt, mark size=2.1pt]
table[x=Eb,y=FER]{figures/data_SCP/256_R12_16_64.txt}; 
\addlegendentry{SCP-64}
\addplot[color=black, mark=x, line width=1pt, mark size=2.1pt]
table[x=Eb,y=FER]{figures/data_SCF/2048_R12_16_8.txt};

\addplot[color=ETSRed!50,   mark=x, line width=1pt, mark size=2.1pt]
table[x=Eb,y=FER]{figures/data_DSCF/2048_R12_16_8.txt};

    \addplot[color=matlab1!100,   mark=x, line width=1pt, mark size=2.1pt]
table[x=Eb,y=FER]{figures/data_SCP/2048_R12_16_8.txt}; 

        \addplot[color=matlab1!100,   mark=o, line width=1pt, mark size=2.1pt]
table[x=Eb,y=FER]{figures/data_SCP/2048_R12_16_64.txt}; 

\draw[dashed,gray!100] 
  (axis cs:2,0.015) ellipse [x radius = 4, y radius = 1,rotate=-45] node [left,black] {}; 
    \node at (axis cs:1.75, 0.0051) [black] {\footnotesize $N=2048$};

    \draw[dashed,gray!100] 
  (axis cs:2.75,0.0175) ellipse [x radius = 5, y radius = 1.5,rotate=-45] node [left,black] {}; 
    \node at (axis cs:3.25, 0.031) [black] {$N=256$};
  \end{semilogyaxis}

\end{tikzpicture}}%\vspace{-6pt}
            \caption{}
            \label{fig:comp_SCF_SCP}
        \end{subfigure}        
        \caption{Error-correction comparison between the bit-flip (SCF-$\Fmax$ \cite{scf_intro} and DSCF-$\Fmax$ \cite{dyn_scf}) algorithms and the elementary perturbation decoders SCP-$\Pmax$ \cite{SCP} for (a) $(1024,496+16)$ and (b) $(256,112+16)$, $(2048,1008+16)$ codes.}        
    \end{figure*}
    
\section{\acrfull{dscfp}}
    In the following, we detail the proposed decoding algorithms combining both flip and perturbation decoders. First, we extend  the error-correction comparison presented in \cite{SCF_SCP_COMP}. % of both decoding algorithms for a same number of additional trials.
    \subsection{Error-correction of Flip and  Perturbation Algorithm}\label{subsec:comparison}
    
    The error-correction comparison carried out in \cite{SCF_SCP_COMP} only focused on \gls{scf} and \gls{scp} for $N\leq512$.
    Next, we extend this comparison with the \gls{dscf} algorithm using a single bit-flip, having better error-correction performance with the same $\Fmax$.
    %Moreover, the \gls{dscf} implementation has a similar area with respect to \gls{scf} since only the metric changes \eqref{eq:metric}.
    The investigated code length is extended to $N\leq 2048$ while the maximum additional trials are set to $\Fmax=\{8,16\}$ and $\Pmax=\{8,16,64\}$.
    Decoders should be compared with the same number of maximum trials $\Tmax$ defined as 
    \begin{align}
        \Tmax=\Fmax+1\,, \,\,\,\quad \Tmax=\Pmax+1\,, \label{eq:tmax}
    \end{align}
    respectively for the flip and the \gls{scp} algorithm.
    For $N=1024$, performances  are shown in \autoref{fig:comp_SCF_SCP_1024}.
    The performance of \gls{dscf} already saturates for $\Fmax=8$ but exhibits the best decoding performance except for \gls{scp} allowing $\Pmax=64$ additional trials.
    \gls{scp} exhibits greater performance for lower \gls{bler}.
    The performance of \gls{scp} does not saturate at $\Pmax=16$ since a gain of $0.25$\,dB is observed at $\text{BLER}=10^{-3}$ with $\Pmax=64$.
    
    \autoref{fig:comp_SCF_SCP} depicts the error-correction performances for $\Fmax=\Pmax=8$ for $N=\{256,2048\}$.
    The performances are similar for \gls{scf} and \gls{scp} while \gls{dscf} beats both decoding algorithms and achieves its ideal error-correction performance.
    %For all decoders, we set the number of additional trials as $\Pmax=\Fmax=\{8,16\}$, which already shows the performance bound for flip decoding algorithm. % except for \gls{scf} and \gls{dscf} where $64$ is avoided because of the saturation.
    The performance of \gls{scp} with $\Pmax=64$ is as well depicted, and exceeds the performance of \gls{dscf} with $\Fmax=8$.
    
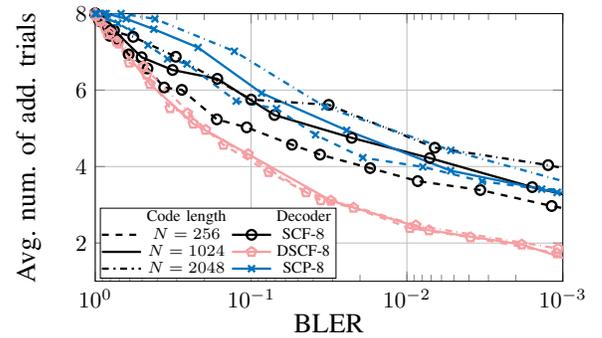
\begin{figure}
        \centering
        \resizebox{0.9\columnwidth}{!}{\begin{tikzpicture}
  \pgfplotsset{
    label style = {font=\fontsize{12pt}{8.2}\selectfont},
    tick label style = {font=\fontsize{10pt}{8.2}\selectfont}
  }
  \begin{semilogxaxis}[
    width=\columnwidth,
    height=0.65\columnwidth,
    xmin=0.001, xmax=1,
    x dir=reverse,
    xlabel={BLER},
    xlabel style={yshift=0.4em},
    ymin=1, ymax=8,
    ylabel style={yshift=-0.6em},
    ylabel={Avg. num. of add. trials},
    yminorticks, xmajorgrids,
    ymajorgrids, yminorgrids,
    %legend pos=outer east,
    legend style={at={(0.01,0.01)},anchor=south west},
    %column font, row sep
    legend style={legend columns=2, font=\scriptsize, row sep=-1mm},
    legend style={fill=white, fill opacity=0.75, draw opacity=1,text opacity=1}, % for future use maybe ? %opacity of filling/border and inside text
    legend style={inner xsep=0.2pt, inner ysep=-1pt}, % TIGHTER
    mark size=1.6pt, mark options=solid,
    ]   

    \addlegendimage{empty legend}
    \addlegendentry{Code length}
        \addlegendimage{empty legend}
    \addlegendentry{Decoder}
    \addplot[line width=1pt,color=black, dashed]       coordinates{    (1,0.1)    };
       \addlegendentry{$N=256$}
           %\addplot[line width=1pt,color=black]       coordinates{    (1,0.1)    };
       %\addlegendentry{$N=512$}
                              \addplot[color=black, mark=o, line width=1pt, mark size=2.1pt]
table[x=FER,y=avg_flip]{figures/data_SCF/1024_R12_16_8.txt}; 
\addlegendentry{SCF-$8$}
           \addplot[line width=1pt,color=black]       coordinates{    (1,0.1)    };
       \addlegendentry{$N=1024$}
       \addplot[color=ETSRed!50,   mark=pentagon, line width=1pt, mark size=2.1pt]
table[x=FER,y=avg_flip]{figures/data_DSCF/1024_R12_16_8.txt}; 
\addlegendentry{DSCF-8}
           \addplot[line width=1pt,color=black,dash dot]       coordinates{    (1,0.1)    };
       \addlegendentry{$N=2048$}

\addplot[color=matlab1!100,   mark=x, line width=1pt, mark size=2.1pt]
table[x=FER,y=avg_perb]{figures/data_SCP/1024_R12_16_8.txt}; 
\addlegendentry{SCP-8}

                \addplot[color=black,dashed, mark=o, line width=1pt, mark size=2.1pt]
table[x=FER,y=avg_flip]{figures/data_SCF/256_R12_16_8.txt};

\addplot[color=ETSRed!50, dashed,  mark=pentagon, line width=1pt, mark size=2.1pt]
table[x=FER,y=avg_flip]{figures/data_DSCF/256_R12_16_8.txt}; 

    \addplot[color=matlab1!100,  dashed, mark=x, line width=1pt, mark size=2.1pt]
table[x=FER,y=avg_perb]{figures/data_SCP/256_R12_16_8.txt};

                \addplot[color=black,dash dot, mark=o, line width=1pt, mark size=2.1pt]
table[x=FER,y=avg_flip]{figures/data_SCF/2048_R12_16_8.txt}; 

\addplot[color=ETSRed!50, dash dot,  mark=pentagon, line width=1pt, mark size=2.1pt]
table[x=FER,y=avg_flip]{figures/data_DSCF/2048_R12_16_8.txt}; 

    \addplot[color=matlab1!100,  dash dot, mark=x, line width=1pt, mark size=2.1pt]
table[x=FER,y=avg_perb]{figures/data_SCP/2048_R12_16_8.txt}; 

  \end{semilogxaxis}

\end{tikzpicture}}%
        \caption{Number of additional trials when required, $\Tmax=9$.}
        \label{fig:comp_trial}
    \end{figure}

    In \cite{SCF_SCP_COMP}, the average number of trials of \gls{scf} and \gls{scp} algorithms are compared.
    Both decoding algorithms tend towards the \gls{sc} complexity since the initial \gls{sc} trial will likely provide a candidate $\hatuvec$ passing the \gls{crc} check for the majority of the frames.
    However, the average number of additional trials when the initial \gls{sc} fails is not provided. 
    It will permit characterizing the effectiveness of the additional trials.
    \autoref{fig:comp_trial} depicts the average number of additional trials when the initial \gls{sc} trial fails in function of the \gls{bler}. 
    For all algorithms, the maximum additional trials  $\Pmax=\Fmax$ is set to 8.
    For all code lengths $N=\{256,1024,2048\}$, \gls{dscf}-8 requires fewer additional trials while \gls{scf} and \gls{scp} require similar additional \gls{sc} iterations.
    Hence, the use of \gls{dscf} decoding algorithm should be prioritized since the effectiveness of its additional trial surpasses the other two.

    %\vspace{-0.021in}
    \subsection{The \gls{dscfp} Decoding Algorithm}
    The performance of \gls{dscf} surpasses \gls{scf} but saturates for low $\Fmax\simeq8$.
    Next, the proposed \gls{dscfp} decoding algorithm is described.
    This decoder aims at improving the performance of \gls{dscf} when the performance saturates. %, at no cost area-wise.
    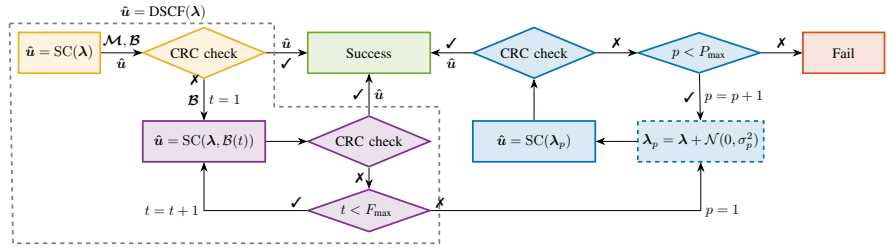
\begin{figure*}[t]
    %\vspace{0.0201in}
        \centering
        \resizebox{0.65\textwidth}{!}{\usetikzlibrary {arrows.meta}
\begin{tikzpicture}
   % Premier rectangle
    \node[very thick, draw=matlab3, minimum width=2cm, minimum height=1cm,fill=matlab3!15] (SC) at (0,0) {$\hatuvec=\SC(\Lvec)$};
    % CRC check
    \node[very thick, draw=matlab3, diamond, minimum width=3cm, minimum height=1cm, align=center,inner sep=-1ex,fill=matlab3!15] (SCCRC) at (SC.east) [anchor=west,xshift=1cm] {CRC check};
    
    \node[very thick, draw=matlab5, minimum width=3cm, minimum height=1cm, fill=matlab5!15] (success) at (SCCRC.east) [anchor=west,xshift=1cm] {Success};
    % outcome
    \draw[arrows = {-Stealth[scale=1.25]}] (SC) -- (SCCRC)  node[midway, above] {$\Mcalbf,\Bcalbf$} node[midway, below] {$\hatuvec$};
    
    \draw[arrows = {-Stealth[scale=1.25]}] (SCCRC) -- (success) node[midway, above] {$\hatuvec$} node[midway,below] {\ding{51}};
    % Deuxième rectangle à droite du losange
    \node[very thick, draw=matlab4,very thick, minimum width=3cm, minimum height=1cm, fill=matlab4!15] (SCF) at (SCCRC.south) [anchor=north,yshift=-1cm] {$\hatuvec=\SC(\Lvec,\Bcal(t))$};
    
    \draw[arrows = {-Stealth[scale=1.25]}] (SCCRC)  -- (SCF) node[midway, left] {$\Bcalbf$} node[midway, right] {$t=1$} node[midway,left,yshift=0.5cm] {\ding{55}};
        % CRC check
    \node[very thick, draw=matlab4, diamond, minimum width=3cm, minimum height=1cm, align=center,inner sep=-1ex,fill=matlab4!15] (SCFCRC) at (SCF.east) [anchor=west,xshift=1cm] {CRC check};
    \draw[arrows = {-Stealth[scale=1.25]}](SCF) -- (SCFCRC);
    \draw[arrows = {-Stealth[scale=1.25]}] (SCFCRC) -- (success) node[midway, right] {$\hatuvec$} node[midway,left] {\ding{51}};
    % Tmax check
   \node[very thick, draw=matlab4, diamond, minimum width=3cm, minimum height=1cm, align=center,inner sep=-1ex,fill=matlab4!15] (Tmax) at (SCFCRC.south) [anchor=north,yshift=-0.5cm] {$t<\Fmax$ };
   
    \draw[arrows = {-Stealth[scale=1.25]}](SCFCRC) -- (Tmax) node [midway,left] {\ding{55}};
   \node[] (southSCF) at (SCF.south)[yshift=-1cm] {};
%    \draw[arrows = {-Stealth[scale=1.25]}] (Tmax.west) node[left,yshift=0.2cm]{\ding{51}}  -- ([yshift=-1.2cm]SCF.south) -- (SCF.south)  node[midway,below,left]{$t=t+1$};
    \draw[arrows = {-Stealth[scale=1.25]}] (Tmax.west) node[left,yshift=0.2cm]{\ding{51}}  -| (SCF.south)  node[midway,below,left]{$t=t+1$};

    \node[very thick, draw=matlab1, minimum width=3cm, minimum height=1cm,fill=matlab1!15] (SCP) at (SCFCRC.east) [anchor=west,xshift=1cm] {$\hatuvec=\SC(\Lpvec)$};
            % CRC check
    \node[very thick, draw=matlab1, diamond, minimum width=3cm, minimum height=1cm, align=center,inner sep=-1ex,fill=matlab1!15] (SCPCRC) at (SCP.north) [anchor=south,yshift=1cm] {CRC check};
    \draw[arrows = {-Stealth[scale=1.25]}] (SCPCRC) -- (success) node [midway, above] {\ding{51}} node [midway, below] {$\hatuvec$};
    % Pmax check
    \node[very thick, draw=matlab1, diamond, minimum width=3cm, minimum height=1cm, align=center,inner sep=-1ex,fill=matlab1!15] (Pmax) at (SCPCRC.east) [anchor=west,xshift=1cm] {$p<\Pmax$};
    \node[very thick, draw=matlab2, minimum width=2cm, minimum height=1cm,fill=matlab2!15] (fail) at (Pmax.east) [anchor=west,xshift=1cm] {Fail};
        \node[very thick, draw=matlab1, dashed, minimum width=3cm, minimum height=1cm,fill=matlab1!15] (perb) at (SCP.east) [anchor=west,xshift=1cm] {$\Lpvec=\Lvec+\mathcal{N}(0,\varp)$};
        %\draw[arrows = {-Stealth[scale=1.25]}] (Tmax.east) node[right,yshift=0.2cm]{\ding{55}}  -- ([yshift=-1.2cm]perb.south) -- (perb.south)  node[midway,below,left]{$p=1$};
        \draw[arrows = {-Stealth[scale=1.25]}] (Tmax.east) node[right,yshift=0.2cm]{\ding{55}}  -| (perb.south)  node[midway,below,right]{$p=1$};
        \draw[arrows = {-Stealth[scale=1.25]}] (perb) -- (SCP);
        
        \draw[arrows = {-Stealth[scale=1.25]}] (SCP) -- (SCPCRC);
        \draw[arrows = {-Stealth[scale=1.25]}] (SCPCRC) -- (Pmax) node[midway, above] {\ding{55}};
        \draw[arrows = {-Stealth[scale=1.25]}] (Pmax) -- (fail) node[midway, above] {\ding{55}};
        \draw[arrows = {-Stealth[scale=1.25]}] (Pmax) -- (perb) node[midway, left] {\ding{51}} node[midway, right] {$p=p+1$};
        
        \draw[-,dashed, gray,very thick] ([yshift=0.2cm,xshift=-0.2cm]SC.north west) -- ([yshift=0.2cm,xshift=-0.8cm]success.north west) --([yshift=-0.8cm,xshift=-0.8cm]success.south west) --([yshift=-0.8cm,xshift=0.2cm]success.south east)-- ([yshift=-0.8cm,xshift=0.2cm]Tmax.east) -- ([yshift=-0.8cm,xshift=-7.2cm]Tmax.west) --([yshift=0.2cm,xshift=-0.2cm]SC.north west);
        \draw ([yshift=0.35cm,xshift=-1cm]SCCRC.north) node[] {$\hatuvec=\text{DSCF}(\Lvec)$};
\end{tikzpicture}}%
        \caption{Block diagram of \gls{dscfp}-$(\Fmax,\Pmax)$ decoding algorithm.}
        \label{fig:diagram_scfp}
    \end{figure*}
    The decoder is described by three parameters $(\Fmax, \Pmax, \varp)$ respectively describing the maximum additional decoding trials including a bit-flip, the maximum additional decoding trials with a corrupted LLR vector, and  the variance of the Gaussian perturbation added to the LLR vector $\Lvec$, creating a corrupted LLR vector $\Lpvec$.
The maximum number of trials $\Tmax$ is then:
\begin{align}
    \Tmax=\Fmax+\Pmax+1\,,\label{eq:tmaxscfp}
\end{align}
where the one corresponds to the initial \gls{sc} trial.
The proposed algorithm is now described. 
First, the initial \gls{sc} algorithm is fed with the LLR vector $\Lvec$ and return $\hatuvec$, i.e., $\hatuvec=\SC(\Lvec)$.
If the CRC is checked, the decoding is considered successful. 
If it fails, the set $\Bcalbf$ of $\Fmax$ bit-flipping candidates that has been computed is used to perform the flipping algorithm.
The set $\Bcalbf$ is designed with the  dynamic metric $\Mcalbf$ \eqref{eq:metric}. %, the decoder is labeled as \acrfull{dscfp} and should be preferred over \gls{scfp}.
Up to $\Fmax$ additional trials are tried, and a CRC check is performed at the end of each trial $t\leq\Fmax$ with the new bit-estimate vector estimated as $\hatuvec=\SC(\Lvec,\Bcal(t))$.
If one of the \gls{crc} checks passes, the decoding is considered successful.
If all checks fail, the LLR vector $\Lvec$ is corrupted independently up to $\Pmax$ times with a noise characterized by  \eqref{eq:noisesimple}.
The corrupted \gls{llr} vector $\Lpvec$ is fed to the \gls{sc} decoder to provide a bit-estimate vector $\hatuvec=\SC(\Lpvec)$.
Up to $\Pmax$ \gls{sc} additional trials are tried, a \gls{crc} check is performed at the end of each trial $p\leq\Pmax$,  the decoding is considered successful if a \gls{crc} check passes.
If all checks fail, the decoding is considered wrong.

\autoref{fig:diagram_scfp} describes the block diagram of the \gls{dscfp} decoding algorithm.
This diagram is composed of 3 distinct parts represented in yellow, purple and blue which respectively correspond to the initial \gls{sc} trial, the flip algorithm (\gls{scf} or \gls{dscf} where the only differences are $\Bcalbf,\Mcalbf$) and the perturbation decoder.
By combining the yellow and the blue parts, the \gls{dscf} algorithm is retrieved and is depicted with a dashed gray area.
The two outcomes of the decoding algorithm, \quot{success} and \quot{fail}, are represented respectively in green and red. 
The decoding can be successful at each of the three parts while the decoding is considered as a failure only after $\Pmax$ trials on the perturbation decoder.

The blue dashed rectangle in \autoref{fig:diagram_scfp}, representing the applied perturbation, is investigated in \cite{dyn_SCP} to enhance the error-correction performance.
The \acrfull{dscp} decoder successively increases $\varp$ if the current trial returns an identical $\hatuvec$.
This comes at the cost of storing all $\hatuvec$, increasing the memory requirements.
While investigating solutions requiring less memory requirements, an approximate solution for not storing all $\hatuvec$ for \gls{dscp} is given next.
The storage of all $\hatuvec$ requires $N\Pmax$ bits, instead of storing $\Pmax$ vectors, the number of ones for each $\hatuvec$ is stored requiring $n\Pmax$ bits. The value of $\varp$ increases if the same number of ones is retrieved.

Finally, solutions to enhanced the basic perturbation \eqref{eq:noisesimple} for the proposed \gls{dscfp} are investigated reusing the locations stored in $\Bcalbf$.
However, no correlation is found between the bit-flip candidates $\Bcalbf$ and the \glspl{llr} in $\Lvec$ that need to have a sign swap when \gls{dscfp} is performing the perturbation part.
Hence, the simple corruption \eqref{eq:noisesimple} will remain for \gls{dscfp}.
\subsection{The \gls{pdscf} Decoding Algorithm}
\begin{figure}[t]
        \centering
        \resizebox{0.85\columnwidth}{!}{\usetikzlibrary {arrows.meta}
\begin{tikzpicture}
    
   % Premier rectangle
    \node[very thick, draw=matlab3, minimum width=3.5cm, minimum height=1cm,fill=matlab3!15] (DSCF) at (0,0) {$\hatuvec=\text{DSCF}(\Lvec)$};

    \node[very thick, draw=matlab5, minimum width=3cm, minimum height=1cm,fill=matlab5!15] (success) at (DSCF.east) [anchor=west,xshift=1cm] {Success};
    \draw[very thick,arrows = {-stealth[scale=2.5]}] (DSCF) -- (success)  node[midway, above] {\ding{51}} node[midway, below] {$\hatuvec$};

    \node[very thick, draw=matlab1, dashed, minimum width=3.5cm, minimum height=1cm,fill=matlab1!15] (perb) at (DSCF.south) [anchor=north,yshift=-1cm] {$\Lpvec=\Lvec+\mathcal{N}(0,\varp)$};
    \draw[very thick,arrows = {-stealth[scale=2.5]}] (DSCF) -- (perb)  node[midway, left] {\ding{55}} node[midway, right] {$p=1$};
    \node[very thick, draw=matlab1, minimum width=3cm, minimum height=1cm,fill=matlab1!15] (PDSCF) at (perb.east) [anchor=west,xshift=1cm] {$\hatuvec=\text{DSCF}(\Lpvec)$};
    \draw[very thick,arrows = {-stealth[scale=2.5]}] (perb) -- (PDSCF);
    \node[very thick, draw=matlab1, diamond, minimum width=3cm, minimum height=1cm, align=center,inner sep=-1ex,fill=matlab1!15] (Pmax) at (PDSCF.east) [anchor=west,xshift=1cm] {$p<\Pmax$};
    
    \draw[very thick,arrows = {-stealth[scale=2.5]}] (PDSCF) -- (Pmax);
    \draw[very thick,arrows = {-stealth[scale=2.5]}] (PDSCF)  -- (success) node[midway,left]{\ding{51}}node[midway,right]{$\hatuvec$};
    \node[very thick, draw=matlab2, minimum width=3cm, minimum height=1cm,fill=matlab2!15] (fail) at (success.east) [anchor=west,xshift=1cm] {Fail};
    %\node[very thick, draw=matlab2, minimum width=2cm, minimum height=1cm] (fail) at (Pmax.north) [anchor=west,yshift=1cm] {Fail};
    \draw[very thick,arrows = {-stealth[scale=2.5]}] (Pmax)  -- (fail) node[midway,left]{\ding{55}};
    \draw[very thick,arrows = {-stealth[scale=2.5]}] (PDSCF)  -- (Pmax) node[midway,above]{\ding{55}};
    \draw[very thick,arrows = {-stealth[scale=2.5]}] (Pmax.south) -- ([yshift=-0.25cm]Pmax.south)-- ([yshift=-0.25cm]perb.south)-- (perb.south);
    \draw[] ([yshift=-0.3cm]PDSCF.south) node[below]{$p=p+1$};
\end{tikzpicture}}%
        \caption{Block diagram of PDSCF-$(\Fmax,\Pmax)$ algorithm.}
        \label{fig:diagram_pscf}
    \end{figure}
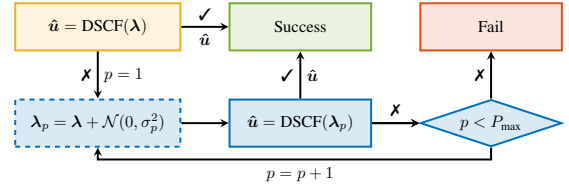

A variant of the \gls{dscfp} algorithm, labeled as \gls{pdscf}, is described next.
The algorithm starts by performing \gls{dscf} with $\Fmax$ as a parameter and $\Lvec$ as input \glspl{llr}.
If the candidate $\hatuvec$ of the initial SC trial $\hatuvec=\SC(\Lvec)$ and all candidates of all $\Fmax$ additional trials $\hatuvec=\SC(\Lvec,\Bcal(t))$ do not check the \gls{crc} code, the vector of \gls{llr} $\Lvec$ is corrupted as $\Lpvec$ \eqref{eq:noisesimple}.
Then, an additional \gls{dscf} iteration with $\Fmax$ trials is performed using the corrupted \glspl{llr} $\Lpvec$.
As usual, a \gls{crc} check is performed on the bit estimates $\hatuvec$ of the initial \gls{scp} trial $\hatuvec=\SC(\Lpvec)$. 
If additional trials are required, the \gls{crc} check is performed on $\hatuvec=\SC(\Lpvec,\Bcal(t))$.
Up to $\Pmax$ additional \gls{dscf} decoders are tried to bound the decoding complexity.
Hence, $\Pmax+1$ \gls{dscf} decoders are potentially tried, each decoder performing up to $\Fmax+1$ trials.
Thus, $\Tmax$ for \gls{pdscf} is expressed as
\begin{align}
    \Tmax=(\Pmax+1)\times(\Fmax+1).\label{eq:tmaxdscf}
\end{align}
In order to compare algorithms, the maximum number of trials $\Tmax$ \eqref{eq:tmax}-\eqref{eq:tmaxscfp}-\eqref{eq:tmaxdscf} should be similar for all decoding algorithms.
The block diagram of the \gls{pdscf} algorithm is depicted in \autoref{fig:diagram_pscf} and reuses the \gls{dscf} block shown in \autoref{fig:diagram_scfp}.
    
    \subsection{Memory Requirements}
\begin{figure}[t]
        \vspace{0.0501in}
            \centering
            \resizebox{0.65\columnwidth}{!}{\begin{tikzpicture}[]

%% SC
\node [draw, fill=yellow!20, rectangle, align=center, inner sep=3pt, minimum height=0.7cm, minimum width=1.3cm] (llr_ch) at (0.3,0) {$\Lvec$ - $\Lpvec$}; 

\node [draw, fill=yellow!20, rectangle, align=center, inner sep=3pt,
minimum height=0.7cm, minimum width=1.3cm] (llr_int) at (0.3,-1.5) {$\Lvec_{\text{int}}$}; 

\node [draw, rectangle, fill=yellow!20, align=center, inner sep=3pt, minimum height=0.7cm, minimum width=1.3cm] (ps) at (2.4,0.0) {$\bm{\hat{u}}$}; 

\node [draw, rectangle, fill=yellow!20, align=center, inner sep=3pt, minimum height=0.7cm, minimum width=1.3cm] (uhat) at (2.4,-1.5) {$\bm{\beta}_{\text{int}}$}; 

\node [draw, fill=matlab4!15, rectangle, align=center, inner sep=3pt, minimum height=0.7cm, minimum width=1.3cm] (metr_flip) at ($(ps.east)+(1.6,0)$) {$\bm{\mathcal{M}}$}; 
\node [draw, fill=matlab4!15, rectangle, align=center, inner sep=3pt, minimum height=0.7cm, minimum width=1.3cm] (bit_flip) at ($(uhat.east)+(1.6,0)$) {$\bm{\mathcal{B}}$};

% \node [draw, fill=red!20, rectangle, align=center, inner sep=3pt, minimum height=0.7cm, minimum width=1.3cm] (ps_rest) at ($(metr_flip.east)+(1.6,0)$) {$\bm{\beta}_{\text{rest}}$}; 

%% Memory lengths
%%% SC
\draw [<->] ($(llr_ch.south)+(-0.6,-0.2)$) -- node[below,font=\footnotesize]{$Q_{\text{ch}}$} ($(llr_ch.south)+(0.6,-0.2)$);
\draw [<->] ($(llr_ch.west)+(-0.2,-0.35)$) -- node[above,rotate=90,font=\footnotesize]{$N$}($(llr_ch.west)+(-0.2,0.35)$);

\draw [<->] ($(llr_int.west)+(-0.2,-0.35)$) -- node[above,rotate=90,font=\footnotesize]{$N-1$}($(llr_int.west)+(-0.2,0.35)$);
\draw [<->] ($(llr_int.south)+(-0.6,-0.2)$) -- node[below,font=\footnotesize]{$Q_{\text{int}}$} ($(llr_int.south)+(0.6,-0.2)$);

\draw [<->] ($(ps.south)+(-0.6,-0.2)$) -- node[below,font=\footnotesize]{$1$} ($(ps.south)+(0.6,-0.2)$);
\draw [<->] ($(ps.west)+(-0.2,-0.35)$) -- node[above,rotate=90,font=\footnotesize]{$N-1$}($(ps.west)+(-0.2,0.35)$);

\draw [<->] ($(uhat.south)+(-0.6,-0.2)$) -- node[below,font=\footnotesize]{$1$} ($(uhat.south)+(0.6,-0.2)$);
\draw [<->] ($(uhat.west)+(-0.2,-0.35)$) -- node[above,rotate=90,font=\footnotesize]{$N$}($(uhat.west)+(-0.2,0.35)$);

%%% SCF
\draw [<->] ($(metr_flip.south)+(-0.6,-0.2)$) -- node[below,font=\footnotesize]{$Q_{\text{flip}}$}($(metr_flip.south)+(0.6,-0.2)$);
\draw [<->] ($(metr_flip.west)+(-0.2,-0.35)$) -- node[above,rotate=90,font=\footnotesize]{$\Fmax$} ($(metr_flip.west)+(-0.2,0.35)$);

\draw [<->] ($(bit_flip.south)+(-0.6,-0.2)$) -- node[below,font=\footnotesize]{$n$} ($(bit_flip.south)+(0.6,-0.2)$);
\draw [<->] ($(bit_flip.west)+(-0.2,-0.38)$) -- node[above,rotate=90,font=\footnotesize]{$F_{\text{max}}$}($(bit_flip.west)+(-0.2,0.38)$);

%%% SCF SRM
%\draw [<->] ($(ps_rest.south)+(-0.64,-0.2)$) -- node[below]{$1$} ($(ps_rest.south)+(0.65,-0.2)$);
%\draw [<->] ($(ps_rest.west)+(-0.2,-0.35)$) -- node[above,rotate=90,font=\footnotesize]{$\nicefrac{N}{2}$}($(ps_rest.west)+(-0.2,0.35)$);
\node [draw, dashed, fill=matlab1!15, rectangle, align=center, inner sep=3pt, minimum height=0.7cm, minimum width=1.3cm] (SCP) at ($(uhat.south)+(-4.5,0.5)$) {$\Lvec$}; 
\draw [<->] ($(SCP.south)+(-0.65,-0.2)$) -- node[below]{$Q_{ch}$} ($(SCP.south)+(0.65,-0.2)$);
\draw [<->] ($(SCP.west)+(-0.2,-0.35)$) -- node[above,rotate=90,font=\footnotesize]{$N$}($(SCP.west)+(-0.2,0.35)$);

%% Dashed borders
%% SC
\node [font=\footnotesize] (sc_lab) at ($(llr_ch.north) + (-1.1,0.22)$) {SC}; 

\draw[very thick] ($(llr_ch.north) + (-1.4,0.4)$) -| ($(uhat.south east)+(0.2,-0.7)$) -| ($(llr_ch.north) + (-1.4,0.4)$);

%SCF
\node [font=\footnotesize, anchor=west] (scf_lab) at ($(metr_flip.north) + (-0.25,0.2)$) {DSCF};

\draw[dashed] ($(llr_ch.north) + (-1.4,0.4)$) -| ($(bit_flip.south east)+(0.2,-0.7)$) -|  ($(llr_ch.north) + (-1.4,0.4)$);

%SCP
\node [font=\footnotesize, anchor=west] (scp_lab) at ($(SCP.base) + (-1.45,2.05)$) {SCP};

\draw[dashed] ($(llr_ch.north) + (-1.4,0.4)$) -| ($(SCP.south west)+(-0.75,-0.8)$) -|  ($(llr_ch.north) + (-1.4,0.4)$);

% SCF-GRM
%\node [font=\footnotesize, anchor=west] (scf_srm_lab) at ($(scf_lab.west) + (0,0.35)$) {SCF with GRM}; 

%\draw[dashed] ($(llr_ch.north) + (-1.9,1.1)$) -- ($(llr_ch.north)+(5.8,1.1)$) -- ($(llr_ch.north)+(5.8,-4.8)$) -- ($(metr_flip.south) + (-1.9,-0.9)$) -- ($(llr_ch.north) + (-1.9,1.1)$);
%\draw[dashed] ($(llr_ch.north) + (-1.7,1.1)$) -| ($(uhat_rest.south east)+(0.2,-1.9)$) -| ($(llr_ch.north) + (-1.7,1.1)$);

\end{tikzpicture}}
            \caption{Memory requirements of \gls{dscfp} and \gls{pdscf}.}
            \label{fig:mem}
        \end{figure}
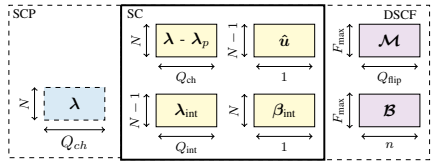

        All aforementioned algorithms are based  on reusing the same single \gls{sc} instance. 
        Next, the memory requirements are given for all algorithms.
        \autoref{fig:mem} depicts the memory for each part of the \gls{dscfp}/\gls{pdscf} algorithms.
        For one block, the vertical arrow depicts the size, while the horizontal depicts the value of the quantization.
        The memory of \gls{sc} is divided in 4 parts, two for the partial sums, and two for the \glspl{llr}.
        The \gls{dscf} algorithm requires the storage of $\Mcalbf$ and $\Bcalbf$.
        The SC instance receives either $\Lvec$ or $\Lpvec$ as inputs.
        The perturbation part requires the storage of the vector $\Lvec$ if $\Pmax>1$ since the perturbation \eqref{eq:noisesimple} is always added to the received \glspl{llr} $\Lvec$.
        If $\Pmax=1$, the perturbation is directly added to the received \glspl{llr}, reducing the memory requirements.
        The memory requirements of \gls{dscfp}/\gls{pdscf} correspond to all memory depicted in \autoref{fig:mem}.
\subsection{Complexity}
%Next, the complexity of the proposed algorithms is evaluated.
The complexity of \gls{sc} is $\mathcal{O}(N\log N)$ \cite{arik_polariz}, all proposed algorithms try up to $\Tmax$ iterations of \gls{sc} \eqref{eq:tmaxscfp}-\eqref{eq:tmaxdscf}.
As a consequence, the complexity of all proposed algorithms is $\mathcal{O}(\Tmax N\log N)$.
However, the average computational complexity at high \gls{snr} is $\mathcal{O}(N\log N)$, since the initial \gls{sc} trial will very likely converge towards the solution which stops the decoding.
% as depicted in \autoref{fig:diagram_scfp}.
%\subsection{Estimation of the Area}

\section{Simulation Results}
    Simulations are carried over the \gls{awgn} with the \gls{bpsk} modulation.
    Codes are designed according to the 5G standard \cite{3GPP_5G_Coding}.
    The notations used are \gls{scf}-$\Fmax$, \gls{dscf}-$\Fmax$, SCP-$\Pmax$, SCL-$L$ while our proposed decoding algorithm are noted \gls{dscfp}-$(\Fmax,\Pmax)$ and \gls{pdscf}-$(\Fmax,\Pmax)$.
    For perturbed algorithms, the value of $\varp$ has a small impact on the performance, $\varp$ is set to $0.95$ as in \cite{SCP}.

\begin{table}[t]
\scriptsize
\centering
\caption{$E_b/N_0$ [dB] to achieve $\text{\gls{bler}}=10^{-3}$, $\frac{K+C}{N}=\frac{1}{2}$} 
\setlength{\tabcolsep}{3pt} % Default value: 6pt
\renewcommand{\arraystretch}{1.3} % Default value: 1
\begin{tabular}{ccccccc}
\toprule
\multicolumn{3}{c}{$N$}& 
$256$ & $512$ & $1024$&$2048$ \\ 
\midrule
dec & Param&$\Tmax$ &  $\nicefrac{E_b}{N_0}$  & $\nicefrac{E_b}{N_0}$  & $\nicefrac{E_b}{N_0}$ & $\nicefrac{E_b}{N_0}$ \\ 
 \midrule
 %\gls{scf} & $13$ & $2.00$ & $19.69$ & $1.875$ & $15.39$ & $2.375$ & $10.87$ \\ 
 \gls{sc}\cite{arik_polariz} & -- & 1 & 4.52& 3.87  & 3.27  & 2.82 \\
 \multirow{2}{*}{\gls{scf}\cite{scf_intro}} & $8$ & 9 & 3.78 & 3.3     &  2.8& 2.52  \\ 
 & $16$ & 17 & 3.55 & 3.2  &     2.7& 2.39  \\ 
 \multirow{2}{*}{\gls{dscf}\cite{dyn_scf}} & $8$ & 9 &3.51    &  3.01 &  $2.59$ &  2.26\\ 
 & $16$ & 17&3.51   &  3.01 &    $2.58$ & 2.25 \\
\multirow{2}{*}{\gls{scp}\cite{SCP}} & $8$ & 9 &  3.87   & 3.28   & 2.75  & 2.45 \\ 
 & $16$ & 17 & 3.65 &   3.01&     2.62& 2.3 \\
 \midrule
 %\gls{scfp} &  $(8,8)$ &  17  & 3.51  & 3.01  & 2.61 &  2.32\\ 
 \gls{dscfp} &  $(8,8)$ & 17   & 3.4  & 2.9  &  2.5&  2.21\\ 
 \gls{pdscf} &  $(7,1)$ & 16   & 3.43  &  2.92 &  2.53& 2.25  \\
\bottomrule
\end{tabular}
\label{tab:recap_fer}
\end{table}
    \subsection{Error-Correction Performance}\label{subsec:sim_perf_ecc}
    In the following, the error-correction of algorithms (SC, SCF/DSCF, SCP, DSCFP, PDSCF) having a similar complexity is investigated. 
    The maximal number of \gls{sc} trials $\Tmax$ is set to 17 for all the decoders.
    The performance of \gls{dscf} saturates for $\Fmax=8$ as discussed in \autoref{subsec:comparison}, i.e., for $\Fmax=16$, same error-correction performance is obtained than with $\Fmax=8$.
    For DSCFP, the parameters are $(\Fmax,\Pmax)=(8,8)$.
    Hence, the performance brought by the flip part is optimal and 8 additional \gls{sc} iterations with corrupted \gls{llr} vector $\Lpvec$ are used to further improved the error-correction performance.
    \autoref{tab:recap_fer} recapitulates all investigated algorithms as well as the parameters and the value of $\Tmax$.
    For \gls{sc}, only one trial is tried, i.e., $\Tmax=1$.
    For existing solutions \cite{scf_intro,dyn_scf,SCP} and proposed algorithms \gls{dscfp} and \gls{pdscf}, results for $\Tmax=\{9,17\}$ are given.
    %For the , results for $\Tmax=17$ are given.
    The \gls{snr}, expressed in $\frac{E_b}{N_0}$, to reach BLER$=0.001$ is given for all decoders and for $N=\{256,512,1024,2048\}$.

\begin{figure}[t]
    %\vspace{0.0501in}
        \centering
        \begin{tikzpicture}
  \pgfplotsset{
    label style = {font=\fontsize{10pt}{8.2}\selectfont},
    tick label style = {font=\fontsize{10pt}{8.2}\selectfont}
  }
  \begin{semilogyaxis}[
    width=\columnwidth,
    height=0.581\columnwidth,
    xmin=1.5, xmax=3,
    xtick={1.5,1.75,...,3},
    xlabel={$E_b/N_0$ [dB]},
    xlabel style={yshift=0.4em},
    ymin=0.00004, ymax=0.12,
    ylabel style={yshift=-0.6em},
    ylabel={BLER},
    yminorticks, xmajorgrids,
    ymajorgrids, yminorgrids,
    legend style={at={(0.03,0.03)},anchor=south west},
    %column font, row sep
    legend style={legend columns=1, font=\small, row sep=-1mm},
    legend style={fill=white, fill opacity=0.7, draw opacity=1,text opacity=1}, % for future use maybe ? %opacity of filling/border and inside text
     legend cell align={left},
    legend style={inner xsep=0.2pt, inner ysep=-1pt}, % TIGHTER
    mark size=1.6pt, mark options=solid,
    ]   

    \addplot[color=matlab3,mark=triangle,  line width=1pt, mark size=2.1pt]
table[x=Eb,y=FER]{figures/data_SC/1024_R12.txt}; 
    \addlegendentry{SC}  
\addplot[color=matlab1,   mark=x, line width=1pt, mark size=2.1pt]
table[x=Eb,y=FER]{figures/data_SCP/1024_R12_16_8.txt}; 
    \addlegendentry{SCP-8}

\addplot[color=matlab1,   mark=square, line width=1pt, mark size=2.1pt]
table[x=Eb,y=FER]{figures/data_SCP/1024_R12_16_16.txt}; 
    \addlegendentry{SCP-16}
    
\addplot[color=ETSRed!50,   mark=x, line width=1pt, mark size=2.1pt]
table[x=Eb,y=FER]{figures/data_DSCF/1024_R12_16_8.txt}; 
    \addlegendentry{DSCF-8}
\addplot[color=ETSRed!50,   mark=square, line width=1pt, mark size=2.1pt]
table[x=Eb,y=FER]{figures/data_DSCF/1024_R12_16_16.txt}; 
    \addlegendentry{DSCF-16}

\addplot[color=matlab5!50,   mark=square*, line width=1pt, mark size=2.1pt]
table[x=Eb,y=FER]{figures/data_DSCFPERB/1024_R12_88.txt}; 
    \addlegendentry{DSCFP $(8,8)$}

    \addplot[color=matlab2,   mark=pentagon, dashed, line width=1pt, mark size=2.1pt]
table[x=Eb,y=FER]{figures/data_PDSCF/1024_R12_28.txt}; 
    \addlegendentry{PDSCF $(7,1)$}

  \end{semilogyaxis}

\end{tikzpicture}
        \caption{Error-correction performance of $(1024,496+16)$ polar codes for decoders requiring a single \gls{sc} instance, $\Tmax\leq17$.}
        \label{fig:perf_1024_perb}
    \end{figure}
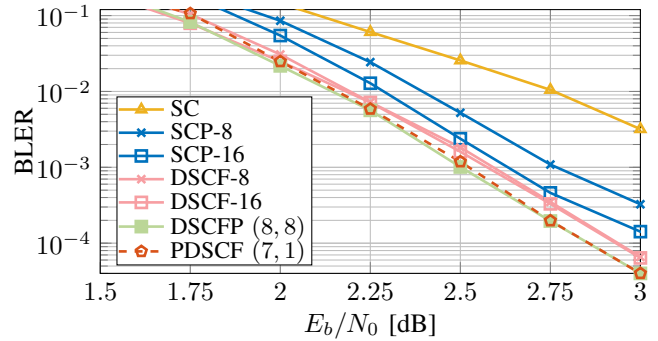

    For all code lengths, the \gls{dscf} performance does not improved from $\Tmax=9$ to $\Tmax=17$ due to the saturation.
    The \gls{dscfp}$-(8,8)$ algorithm has better performance with respect to all decoding algorithms.
    %It is due to the more accurate metric used by \gls{dscf} with respect to \gls{scf}.
    By adding $8$ perturbed \gls{sc} decoding, the performance is improved by around $0.1$\,dB.
    Note that the performance tends to reduce with greater code length $N$.
    The \gls{pdscf}$-(7,1)$ algorithm exhibits a slight loss with respect to \gls{dscfp}, but perform one trial less.
    However, all proposed algorithms have better performance with respect to the existing solutions \gls{scf}, \gls{dscf} or \gls{scp} for same number of trials $\Tmax=17$.
    \autoref{fig:perf_1024_perb} depicts the decoding performance of the proposed algorithms for $(1024,496+16)$ and of \gls{dscf}-8, \gls{dscf}-16, \gls{scp}-8, and \gls{scp}-16.
    The performance saturates for \gls{dscf} while the additional perturbed trials for the proposed algorithms allow to overcome this bound.
    %The performance of \gls{scp}-8 is shown as well for reference.
    
\subsection{Influence of the Maximum Number of Trials $\Tmax$}
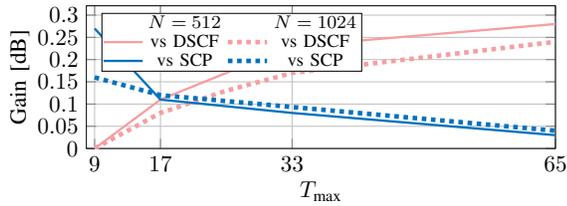
\begin{figure}[t]
    %%\vspace{0.021in}
        \centering
        \resizebox{0.88\columnwidth}{!}{\begin{tikzpicture}
  \pgfplotsset{
    label style = {font=\fontsize{10pt}{8.2}\selectfont},
    tick label style = {font=\fontsize{10pt}{8.2}\selectfont}
  }
  \begin{axis}[
    width=\columnwidth,
    height=0.43\columnwidth,
    xmin=8, xmax=65,
    xtick={9,17,33,65},
    xlabel={$\Tmax$},
    xlabel style={yshift=0.4em},
    ymin=0, ymax=0.321,
    ytick={0,0.05,0.1,...,0.35},
    yticklabels={{$0$},{$0.05$},{$0.1$},{$0.15$},{$0.2$},{$0.25$},{$0.3$},{$0.35$}},
    ylabel style={yshift=-0.6em},
    ylabel={Gain [dB]},
    yminorticks, xmajorgrids,
    ymajorgrids, yminorgrids,
    legend style={at={(0.31,0.95)},anchor=north},
    %column font, row sep
    legend style={legend columns=2, font=\footnotesize, row sep=-1mm},
    legend style={fill=white, fill opacity=0.7, draw opacity=1,text opacity=1}, % for future use maybe ? %opacity of filling/border and inside text
    legend style={inner xsep=0.2pt, inner ysep=-1pt}, % TIGHTER
    mark size=1.6pt, mark options=solid,
    ]   
        \addlegendimage{empty legend}
    \addlegendentry{$N=512$}
        \addlegendimage{empty legend}
    \addlegendentry{$N=1024$}

    \addplot[line width=1pt,color=ETSRed!50] table[row sep=crcr]{%
                9	0\\
                17 0.11\\
                33 0.23\\
                65 0.28\\
                };
    \addlegendentry{vs DSCF}
    \addplot[line width=2pt,color=ETSRed!50,dotted] table[row sep=crcr]{%
                9	0\\
                17 0.08\\
                25 0.13\\
                33 0.17\\
                65 0.24\\
    };
    \addlegendentry{vs DSCF}
    
    \addplot[line width=1pt,color=matlab1] table[row sep=crcr]{%
                9	0.27\\
                17 0.11\\  
                33 0.08\\
                65 0.03\\
    };
    
    \addlegendentry{vs SCP}
    \addplot[line width=2pt,color=matlab1, dotted] table[row sep=crcr]{%
                9	0.16\\
                17 0.12\\  
                65 0.04\\
    };
    \addlegendentry{vs SCP}
       %\addlegendimage{only marks, mark=diamond*, mark options={scale=1.8, fill=blue}}

  \end{axis}

\end{tikzpicture}}%%\vspace{-15pt}
        \caption{Gain of DSCFP $(8,\Tmax-8)$ over DSCF and SCP at $\text{BLER}=10^{-3}$ for various values of $\Tmax$, $\frac{K+C}{N}=1024$.}
        \label{fig:gain}
    \end{figure}

    Next, the influence of $\Tmax$ is described for the proposed algorithms.
    $\Fmax$ is still set to $8$ for our proposed algorithm due to the saturation.
    \autoref{fig:gain} depicts the gain of \gls{dscfp} with respect to \gls{dscf} and \gls{scp} for various values of $\Tmax$, $N=\{512,1024\}$, and the code rate $\nicefrac{1}{2}$.
    An ascending gain is expected with respect to \gls{dscf} since its performance saturates for $\Fmax=8$ ($\Tmax=9$).
    The gain with respect to \gls{scp} corresponds to the gain by replacing $\Fmax=8$ trials with bit-flipping with a \gls{sc} additional trial on a perturbed \gls{llr} vector $\Lpvec$.
    The gain is expected to be higher for small $\Tmax$. % while reducing with greater $\Tmax$.
    
    For $\Tmax=9$, the proposed \gls{dscfp} algorithm is equivalent to \gls{dscf}, hence there is no gain with respect to \gls{dscf} while the gain is maximized with respect to \gls{scp} as depicted in \autoref{subsec:comparison}.
    For $\Tmax=17$, the gain of around $0.1$\,dB described in \autoref{subsec:sim_perf_ecc} is retrieved with respect to both decoding algorithms.
    For $\Tmax=64$, the gain is of $0.25$\,dB and  $0.04$\,dB with respect to \gls{dscf} and \gls{scp}.
    
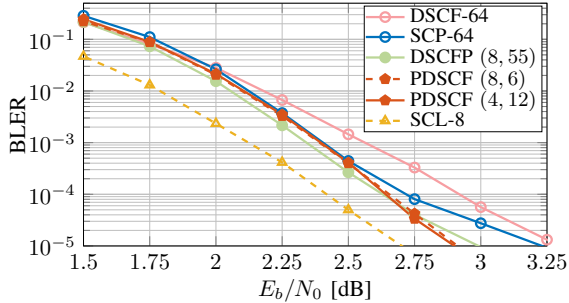
\begin{figure}[t]
        \centering
        \resizebox{0.88\columnwidth}{!}{\begin{tikzpicture}
  \pgfplotsset{
    label style = {font=\fontsize{10pt}{8.2}\selectfont},
    tick label style = {font=\fontsize{10pt}{8.2}\selectfont}
  }
  \begin{semilogyaxis}[
    width=\columnwidth,
    height=0.61\columnwidth,
    xmin=1.5, xmax=3.25,
    xtick={1.5,1.75,2.0,...,3.75},
    xlabel={$E_b/N_0$ [dB]},
    xlabel style={yshift=0.4em},
    ymin=0.00001, ymax=0.5,
    ylabel style={yshift=-0.6em},
    ylabel={BLER},
    yminorticks, xmajorgrids,
    ymajorgrids, yminorgrids,
    legend style={at={(0.99,0.99)},anchor=north east},
    %column font, row sep
    legend style={legend columns=1, font=\small, row sep=-1mm},
    legend style={fill=white, fill opacity=0.85, draw opacity=1,text opacity=1}, % for future use maybe ? %opacity of filling/border and inside text
     legend cell align={left},
    legend style={inner xsep=0.2pt, inner ysep=-1pt}, % TIGHTER
    mark size=1.6pt, mark options=solid,
    ]   

        \addplot[color=ETSRed!50,  mark=o, line width=1pt, mark size=2.1pt]
table[x=Eb,y=FER]{figures/data_DSCF/1024_R12_16_64.txt}; 
    \addlegendentry{DSCF-64}
        \addplot[color=matlab1,  mark=o, line width=1pt, mark size=2.1pt]
table[x=Eb,y=FER]{figures/data_SCP/1024_R12_16_64.txt}; 
    \addlegendentry{SCP-64}
        \addplot[color=matlab5!50, mark=*,  line width=1pt, mark size=2.1pt]
table[x=Eb,y=FER]{figures/data_DSCFPERB/1024_R12_8_55.txt}; 
    \addlegendentry{DSCFP $(8,55)$}
    \addplot[color=matlab2, dashed, mark=pentagon*,line width=1pt, mark size=2.1pt]
table[x=Eb,y=FER]{figures/data_PDSCF/1024_R12_68.txt}; 
    \addlegendentry{PDSCF $(8,6)$}
    
    \addplot[color=matlab2, mark=pentagon*, line width=1pt, mark size=2.1pt]
table[x=Eb,y=FER]{figures/data_PDSCF/1024_R12_12_4.txt}; 
    \addlegendentry{PDSCF $(4,12)$}
        \addplot[color=matlab3,mark=triangle, dashed,  line width=1pt, mark size=2.1pt]
table[x=Eb,y=FER]{figures/data_SCL/1024_R12.txt}; 
    \addlegendentry{SCL-8}  
  \end{semilogyaxis}

\end{tikzpicture}}%%\vspace{-15pt}
        \caption{$(1024,496+16)$ BLER performance, $\Tmax=64$.}
        %Error-correction performance for $\Tmax=64$, $N=1024,\frac{K+C}{N}=\frac{1}{2}$.}
        \label{fig:perf_pdscf}
    \end{figure}

\autoref{fig:perf_pdscf} depicts the performance of the proposed algorithms for $\Tmax\simeq65$.
The depicted performances are for \gls{dscfp}-$(8,56)$ ($\Tmax=65$ \eqref{eq:tmaxscfp}), \gls{pdscf}-$(6,8)$ ($\Tmax=63$ \eqref{eq:tmaxdscf}), and \gls{pdscf}-$(12,4)$ ($\Tmax=65$ \eqref{eq:tmaxdscf}).
For reference, the error-correction performances of \gls{scp}-64 \cite{SCP}, \gls{dscf}-64 \cite{dyn_scf}, and \gls{scl}-8 are shown.
\gls{scp} has greater performance with respect to \gls{dscf}, however the gain diminishes for $\text{BLER}<10^{-4}$ with a crossover at $\text{BLER}=3\cdot10^{-6}$.
For $\text{BLER}>10^{-4}$, the proposed \gls{dscfp} improves the error-correction performance of \gls{scp} with a gain of $\simeq0.08$\,dB.
For lower \gls{bler}, the presence of \gls{dscf} permits to increase the gap.
At $\text{BLER}=10^{-6}$, the gain is of $0.25$ and $>0.375$\,dB with respect to \gls{dscf}-64 and \gls{scp}-64.
The two performances of \gls{pdscf} are almost equivalent and has a similar behavior regardless of the \gls{bler}.
For $\text{BLER}=4\cdot10^{-5}$, \gls{dscfp} has greater performance. % due to the presence of more perturbed \gls{sc}. %which permits greater performance according to the performance of \gls{dscf} and \gls{scp}.
However, \gls{pdscf} allows for a greater error-correction after.
At $\text{BLER}=10^{-6}$, the gains of \gls{pdscf} with respect to \gls{dscfp}, \gls{dscf}, and \gls{scp} are of $0.125$, $0.375$, $>0.5$\,dB, respectively. 
\gls{scl}-8, having a more complex implementation, shows the best performance. with $0.27$ dB gain at $\text{BLER}=10^{-2}$ and $0.2$ dB gain at $\text{BLER}=10^{-6}$.

    \subsection{Average Computational Complexity}
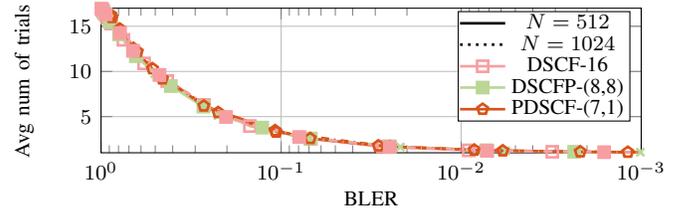
\begin{figure}[t]
    %%\vspace{0.021in}
        \centering
        \begin{tikzpicture}
  \pgfplotsset{
    label style = {font=\fontsize{8pt}{8.2}\selectfont},
    tick label style = {font=\fontsize{8pt}{8.2}\selectfont}
  }
  \begin{semilogxaxis}[
    width=\columnwidth,
    height=0.4\columnwidth,
    xmin=0.001, xmax=1,
    x dir=reverse,
    xlabel={BLER},
    xlabel style={yshift=0.4em},
    ymin=1, ymax=17,
    ylabel style={yshift=-0.6em},
    ylabel={Avg num of trials},
    yminorticks, xmajorgrids,
    ymajorgrids, yminorgrids,
    %legend pos=outer east,
    legend style={at={(0.98,0.98)},anchor=north east},
    %column font, row sep
    legend style={legend columns=1, font=\footnotesize, row sep=-1mm},
    legend style={fill=white, fill opacity=1, draw opacity=1,text opacity=1}, % for future use maybe ? %opacity of filling/border and inside text
    legend style={inner xsep=0.2pt, inner ysep=-1pt}, % TIGHTER
    mark size=1.6pt, mark options=solid,
    ]   
    
           \addplot[line width=1pt,color=black]       coordinates{    (1,0.1)    };
       %\addlegendimage{only marks, mark=diamond*, mark options={scale=1.8, fill=blue}}
       \addlegendentry{$N=512$}
       
           \addplot[line width=1pt,color=black,dotted]       coordinates{    (1,0.1)    };
       %\addlegendimage{only marks, mark=diamond*, mark options={scale=1.8, fill=blue}}
       \addlegendentry{$N=1024$}
%            \addplot[color=matlab1, mark=square, line width=1pt, mark size=2.1pt]
%table[x=FER,y=avg_all]{figures/data_SCFPERB/512_R12_88.txt}; 
%    \addlegendentry{SCFP-(8,8)}
                \addplot[color=ETSRed!50, mark=square, line width=1pt, mark size=2.1pt]
table[x=FER,y=avg_all]{figures/data_DSCF/512_R12_16_16.txt}; 
    \addlegendentry{DSCF-16}
            \addplot[color=matlab5!50, mark=square*, line width=1pt, mark size=2.1pt]
table[x=FER,y=avg_all]{figures/data_DSCFPERB/512_R12_88.txt}; 
    \addlegendentry{DSCFP-(8,8)}

                \addplot[color=matlab2, mark=pentagon, line width=1pt, mark size=2.1pt]
table[x=FER,y=avg_all]{figures/data_PDSCF/512_R12_28.txt}; 
    \addlegendentry{PDSCF-(7,1)}

            \addplot[color=matlab5!50, dash dot,mark=x, line width=1pt, mark size=2.1pt]
table[x=FER,y=avg_all]{figures/data_DSCFPERB/1024_R12_88.txt}; 
  
                  \addplot[color=matlab2, mark=pentagon,dash dot, line width=1pt, mark size=2.1pt]
table[x=FER,y=avg_all]{figures/data_PDSCF/1024_R12_28.txt}; 
                \addplot[color=ETSRed!50, dash dot,mark=square*, line width=1pt, mark size=2.1pt]
table[x=FER,y=avg_all]{figures/data_DSCF/1024_R12_16_16.txt}; 
    \end{semilogxaxis}

\end{tikzpicture}%%\vspace{-15pt}
        \caption{Average computational complexity of proposed algorithms for $N=\{512,1024\}$ and rate $R=\frac{K+C}{N}=\frac{1}{2}$.}
        \label{fig:comp_perb}
    \end{figure}

    \autoref{fig:comp_perb} depicts the average number of trials required, an estimation of the average computational complexity of the proposed algorithms. 
    Throughout the practical region of \gls{bler}, results indicate that the average number of trials is approximately the same for the proposed algorithms \gls{dscfp} and \gls{pdscf} with respect to \gls{dscf}. 
    Meanwhile, the proposed \gls{dscfp} offers a coding gain of $0.1$\,dB over \gls{dscf}. 
    
\section{Conclusion}
    This paper proposes two decoding algorithms, DSCFP and PDSCF, for CRC-aided polar codes. The goal is to enhance the error-correction performance of DSCF while achieving the lowest possible implementation complexity.
    In order to restrain the implementation complexity, the single bit-flip \gls{dscf} is first tried and if it fails, a pre-defined Gaussian perturbation is applied.
    The proposed algorithms match or surpass the performance of existing algorithms with similar implementation complexity.
    At $\text{BLER}=10^{-3}$, code rate $\nicefrac{1}{2}$, and $N=\{512,1024\}$, \gls{dscfp} has similar performance with respect to \gls{dscf} for $\Tmax=9$ and $0.1$\,dB gain  for $\Tmax=17$.
    For $\Tmax=64$, the gain is of $0.2$\,dB at $\text{BLER}=10^{-3}$ and of $0.375$\,dB at $\text{BLER}=10^{-6}$ for \gls{pdscf} while still having an average computational complexity similar to that of \gls{sc}.

\bibliographystyle{IEEEtran}
\bibliography{IEEEabrv,ConfAbrv,references}
\end{document}